\documentclass[onecolumn,aps,pra,floatfix,twocolumn]{revtex4-2}
\usepackage{graphicx}
\usepackage{times}
\usepackage{nicefrac}
\usepackage{amsmath}
\usepackage{amsfonts}
\usepackage{amssymb}
\usepackage{amsthm}
\usepackage{epsf}
\usepackage{bm}
\usepackage{bbm}
\usepackage{times}

\usepackage{dcolumn}

\newcolumntype{.}{D{x}{}{-1}}
\newcolumntype{w}[1]{D{.}{.}{#1}}

\begin{document}

\newcommand{\half}{\frac12}
\newcommand{\vare}{\varepsilon}
\newcommand{\eps}{\epsilon}
\newcommand{\pr}{^{\prime}}
\newcommand{\ppr}{^{\prime\prime}}
\newcommand{\pp}{{p^{\prime}}}
\newcommand{\ppp}{{p^{\prime\prime}}}
\newcommand{\hp}{\hat{\bfp}}
\newcommand{\hr}{\hat{\bfr}}
\newcommand{\hk}{\hat{\bfk}}
\newcommand{\hx}{\hat{\bfx}}
\newcommand{\hpp}{\hat{\bfpp}}
\newcommand{\hq}{\hat{\bfq}}
\newcommand{\rqq}{{\rm q}}
\newcommand{\bfk}{{\bm{k}}}
\newcommand{\bfp}{{\bm{p}}}
\newcommand{\bfq}{{\bm{q}}}
\newcommand{\bfr}{{\bm{r}}}
\newcommand{\bfx}{{\bm{x}}}
\newcommand{\bfy}{{\bm{y}}}
\newcommand{\bfz}{{\bm{z}}}
\newcommand{\bfpp}{{\bm{\pp}}}
\newcommand{\bfppp}{{\bm{\ppp}}}
\newcommand{\balpha}{\bm{\alpha}}
\newcommand{\bvare}{\bm{\vare}}
\newcommand{\bgamma}{\bm{\gamma}}
\newcommand{\bGamma}{\bm{\Gamma}}
\newcommand{\bLambda}{\bm{\Lambda}}
\newcommand{\bmu}{\bm{\mu}}
\newcommand{\bnabla}{\bm{\nabla}}
\newcommand{\bvarrho}{\bm{\varrho}}
\newcommand{\bsigma}{\bm{\sigma}}
\newcommand{\bTheta}{\bm{\Theta}}
\newcommand{\bphi}{\bm{\phi}}
\newcommand{\bomega}{\bm{\omega}}
\newcommand{\intzo}{\int_0^1}
\newcommand{\intinf}{\int^{\infty}_{-\infty}}
\newcommand{\lbr}{\langle}
\newcommand{\rbr}{\rangle}
\newcommand{\ThreeJ}[6]{
        \left(
        \begin{array}{ccc}
        #1  & #2  & #3 \\
        #4  & #5  & #6 \\
        \end{array}
        \right)
        }
\newcommand{\SixJ}[6]{
        \left\{
        \begin{array}{ccc}
        #1  & #2  & #3 \\
        #4  & #5  & #6 \\
        \end{array}
        \right\}
        }
\newcommand{\NineJ}[9]{
        \left\{
        \begin{array}{ccc}
        #1  & #2  & #3 \\
        #4  & #5  & #6 \\
        #7  & #8  & #9 \\
        \end{array}
        \right\}
        }
\newcommand{\Vector}[2]{
        \left(
        \begin{array}{c}
        #1     \\
        #2     \\
        \end{array}
        \right)
        }

\newcommand{\Dmatrix}[4]{
        \left(
        \begin{array}{cc}
        #1  & #2   \\
        #3  & #4   \\
        \end{array}
        \right)
        }
\newcommand{\Dcase}[4]{
        \left\{
        \begin{array}{cl}
        #1  & #2   \\
        #3  & #4   \\
        \end{array}
        \right.
        }
\newcommand{\cross}[1]{#1\!\!\!/}

\newcommand{\Za}{{Z \alpha}}
\newcommand{\im}{{ i}}


\title{Relativistic Bethe logarithm for triplet states of helium-like ions}

\author{Vladimir A. Yerokhin}
\affiliation{Peter the Great St.~Petersburg Polytechnic University,
Polytekhnicheskaya 29, 195251 St.~Petersburg, Russia}

\author{Vojt\v{e}ch Patk\'o\v{s}}
\affiliation{Faculty of Mathematics and Physics, Charles University,  Ke Karlovu 3, 121 16 Prague 2, Czech Republic}

\author{Krzysztof Pachucki}
\affiliation{Faculty of Physics, University of Warsaw,
             Pasteura 5, 02-093 Warsaw, Poland}

\date{\today}

\begin{abstract}

We report a calculation of relativistic corrections of order $m\alpha^7$ to the Bethe logarithm for the $2\,^3S$
and $2\,^3P$ states of helium-like ions. The calculation is required for improving the accuracy of theoretical
energies of helium-like ions and for checking the evaluation of the $m\alpha^7$ effects in helium
performed in [V.~Patk\'o\v{s}, V.~A.~Yerokhin, K.~Pachucki, Phys.~Rev.~A {\bf 103}, 042809 (2021)],
where a significant discrepancy with experimental results was found. 
The large-$Z$ limit of the relativistic Bethe logarithm
is determined numerically, in excellent agreement with the analytical results obtained from the hydrogen theory.

\end{abstract}

\maketitle

\section{Introduction}

The dominant contribution to the Lamb shift of an atomic energy level is induced by the electron self-energy. The nonrelativistic
part of it was first described by Bethe \cite{bethe:47} in terms of the logarithm of the mean excitation energy, which is nowadays
called the Bethe logarithm. The Bethe logarithm involves a summation over the complete spectrum of the
Schr\"odinger equation, which is nearly divergent because of large contributions from high-energy continuum states.
The calculation of the  Bethe logarithm is a relatively straightforward task in the case of
hydrogen, because the electron propagator
is known analytically. For atoms with more than one electron, the task becomes more challenging.
Accurate calculations of
the Bethe logarithm for the helium atom have long been considered to be a difficult problem but are presently
well established \cite{drake:99:cjp,korobov:99}. The most accurate results for helium were obtained
in Ref.~\cite{korobov:19:bethelog} and for helium-like ions in Refs.~\cite{drake:99:cjp,yerokhin:10:helike}.

The Bethe logarithm is a part of the leading QED correction that is of order $m\alpha^5$ for light atoms (where $m$ is the
electron mass and $\alpha$ is the fine-structure constant). At the present level of experimental and theoretical interest,
QED effects of higher orders in $\alpha$ need to be accounted for. One of the dominant effects of order $m\alpha^7$
is the relativistic
correction to the Bethe logarithm. It appeared first in the hydrogen theory, where it was evaluated in Refs.~\cite{pachucki:93,jentschura:96,jentschura:03:prl}.
Later these calculations were extended to the two-center problem \cite{korobov:13}.
For the helium atom, the relativistic Bethe logarithm was calculated for the fine-structure
\cite{pachucki:06:prl:he,pachucki:09:hefs,pachucki:10:hefs} and recently for the $2\,^3S$ and $2\,^3P$ states
\cite{yerokhin:18:betherel}. In the present work, we improve the numerical accuracy
for the helium atom and extend calculations to helium-like
ions.

This work is in part motivated by the recent observation of a
significant discrepancy between
theoretical predictions and experimental results for the ionization energies of the triplet $n = 2$ states
in the helium atom \cite{patkos:21:helamb,clausen:21}. In view of this discrepancy, it is important to
cross-check the calculations of the $m\alpha^7$ effects reported in
Refs.~\cite{patkos:21:helamb,yerokhin:18:betherel,patkos:20}. A way to
check calculations for helium is to perform analogous computations for helium-like ions
with different values of the nuclear charge number $Z$ and, by fitting the $1/Z$ expansion,
determine the large-$Z$ limit of the corresponding corrections. This limit
should agree with analytical results obtained from the hydrogen theory.

The goal of the present work is to compute the relativistic correction to the Bethe logarithm
for the $2\,^3S$ and $2\,^3P$ states of helium-like atoms with $Z = 2\,$--$\,12$. By studying the
$Z$ dependence of the numerical results we will determine their high-$Z$ limit and compare it with
the values obtained from the hydrogen theory. This cross-check will test the consistency of
the helium calculations with the more established calculations for hydrogen. In addition, the
obtained results for the relativistic Bethe logarithm will be later used to improve the accuracy of
theoretical predictions for the energy levels of helium-like ions.

%
\section{Basic formulas}

\subsection{Nonrelativistic Bethe logarithm}

The starting point of the theoretical description
is the nonrelativistic Hamiltonian for an atom in the presence of external electromagnetic fields,
\begin{align}\label{eq:1}
{\cal H} =&\,\sum_a \frac{\vec \pi_a^2}{2\,m} + V + e\,\sum_a \phi(\vec r_a)
\end{align}
where $\vec\pi_a = \vec p_a-e\,\vec A(\vec r_a)$,  $\phi(\vec r_a)$ and $\vec A(\vec r_a)$
are the external scalar and vector potentials, respectively,
\begin{align}\label{eq:1b}
V = -\sum_a\frac{\Za}{r_a}  + \sum_{a < b}\frac{\alpha}{r_{ab}}\,,
\end{align}
and the summation indices $a$ and $b$ run over the electrons in the atom.

The nonrelativistic low-energy part of the one-loop electron self-energy 
is obtained from the Hamiltonan (\ref{eq:1}) and has the form
\begin{align}\label{eq:2}
E_{L}(\Lambda) &\ = \frac{e^2}{m^2} \int_{|\bfk| < \Lambda} \frac{d^3k}{(2\pi)^3 2 k}
    \left( \delta^{ij} - \hk^i\hk^j\right)\,
    \nonumber \\ & \times
    \sum_{ab}\lbr \psi| p^i_a \, e^{i\bfk\cdot\bfr_a} \, \frac1{E-H-k} \, p^j_b \, e^{-i\bfk\cdot\bfr_b}|\psi\rbr
    \,,
\end{align}
where $\hk = \bfk/k$, $\Lambda$ is the high-momentum cutoff parameter,
$H$ and $E$ are the nonrelativistic Hamiltonian 
(without the external electromagnetic field) and its eigenvalue, respectively.
To the leading order in $\alpha$,
the exponential factors $e^{i\bfk\cdot\bfr}$ can be neglected.
Performing the integration over $\hk$,
we arrive at known formulas for the low-energy contribution of order $m\alpha^5$\,,
\begin{align}\label{eq:3}
E_{L}^{(5)}(\Lambda) &\ =  \frac{2\alpha}{3\pi m^2} \int_0^{\Lambda} dk\,k\,P_{nd}(k)\,,
 \\
P_{nd}(k) &\ = \Big< \vec{P} \, \frac1{E-H-k} \, \vec{P} \Big> \,,
\end{align}
where $\vec{P} \equiv \sum_a\vec{p}_a $.

Since $E_{L}(\Lambda)$ diverges as $\Lambda \to \infty$, one obtains the finite 
result by subtracting the divergent terms of the large-$\Lambda$ asymptotics 
and then performing the limit $\Lambda \to \infty$.
The large-$k$ expansion of $P_{nd}(k)$ reads as
\begin{align}\label{eq:4}
k\, P_{nd}(k) = \left< \nabla^2 \right> + \frac1{k} \, D + \ldots\,,
\end{align}
where $\vec{\nabla}\equiv \sum_a \vec{\nabla}_a$ and $D = 2\pi Z\lbr \sum_a\delta^3(r_a)\rbr$.

The $m\alpha^5$ low-energy contribution is standardly expressed in terms of the Bethe logarithm
$\ln k_0$, which represents the finite part of Eq.~(\ref{eq:3}) as
\begin{align}\label{eq:5}
\ln k_0 =&\ \frac{\langle \vec P\,(H-E)\,\ln[2\,(H-E)/E_h]\,\vec P\rangle}{\langle \vec P\,(H-E)\,\vec P\rangle}\nonumber \\
=&\  - \frac1{D} \int_0^{\infty} \! dk \biggl[ k\, P_{nd}(k)
 - \left< \nabla^2 \right> - \frac{D}{k}\, \theta(k-E_h/2)\biggr]\,,
\end{align}
where $\theta(x)$ is the Heaviside $\theta$ function,
$\theta(x) = 0$ for $x <0$ and 1 for $x >=0$, and $E_h = m\alpha^2$ is the Hartree energy.

\subsection{Relativistic Bethe logarithm}

In the present work we are interested in the relativistic corrections to the Bethe logarithm.
They can be obtained from the Breit Hamiltonian in the presence of external electromagnetic fields.
Since we are interested in the center-of-gravity
energy levels, it is sufficient to take into account only the
spin-independent part of the Breit Hamiltonian. It is given by
\begin{align}\label{eq:7}
{\cal H}^{(4)}_{\rm Breit} &\ = \sum_a \bigg[-\frac{\pi_a^4}{8m^3}+
\frac{\pi Z\alpha}{2m^2}\,\delta^3(r_a)\bigg]
\nonumber\\&
+ \sum_{a < b} \bigg[ \frac{\pi\alpha}{m^2}\,\delta^3(r_{ab})
-\frac{\alpha}{2m^2}\,\pi_a^i\,
\biggl(\frac{\delta^{ij}}{r_{ab}}+\frac{r^i_{ab}\,r^j_{ab}}{r_{ab}^3}\biggr)\,\pi_{b}^j
\bigg]\,.
\end{align}
From this Hamiltonian we obtain the relativistic correction to the Bethe logarithm of order
$m\alpha^7$ as a sum of three parts,
\begin{align} \label{eq:11}
E_L^{(7)}(\Lambda)&\ =  E_{L1}(\Lambda) + E_{L2}(\Lambda) + E_{L3}(\Lambda)
\nonumber \\ &
 = \frac{2\alpha}{3\pi m^2} \int_0^{\Lambda}dk\, k \Big[ P_{L1}(k) + P_{L2}(k) + P_{L3}(k) \Big]\,.
\end{align}
The first part is a perturbation of the $m\alpha^5$ contribution by the Breit Hamiltonian 
(without external electromagnetic fields),
\begin{align} \label{eq:12}
P_{L1}(k) &\ = 2 \left< H_{\rm Breit} \frac1{(E-H)'}\, \vec{P}\,  \frac1{E-H-k} \, \vec{P} \right> \,,
\nonumber \\ & +
\left< \vec{P}\,  \frac1{E-H-k} \,\bigl[ H_{\rm Breit} - \lbr H_{\rm Breit}\rbr\bigr]  \frac1{E-H-k} \,\vec{P} \right> \,.
\end{align}
The second part is induced by the correction to the current,
\begin{align} \label{eq:13}
P_{L2}(k) &\ = 2\, \left< \vec{\delta j} \, \frac1{E-H-k} \, \vec{P} \right> \,.
\end{align}
The correction to the current $\delta j^i$ is obtained from the Breit-Pauli Hamiltonian,
specifically from the first and fourth terms of Eq.~(\ref{eq:7}), with the result
\begin{align} \label{eq:10}
\delta j^i = -\frac1{2m^2}\,\sum_a p_a^i\,p_a^2 -\frac{\alpha}{2m}
   \sum_{a,b}\bigg(\frac{\delta^{ij}}{r_{ab}} + \frac{r_{ab}^ir_{ab}^j}{r_{ab}^3}\bigg)\, p_b^j\,.
\end{align}
Finally, the third part is the retardation correction induced by the expansion of the exponential functions
in Eq.~(\ref{eq:2}),
\begin{widetext}
\begin{align} \label{eq:14}
P_{L3}(k) &\ = \frac{3k^2}{8\pi} \int d\hk
 \left( \delta^{ij} - \hk^i\hk^j\right)\,
 \biggl[
\Big< \sum_a p_a^i\, (\hk\cdot{\bm{r_a}})\, \frac1{E-H-k} \,\sum_b (\hk\cdot\bfr_b) p_b^j \Big>
-
\Big< \sum_a p_a^i\, (\hk\cdot\bfr_a)^2\, \frac1{E-H-k} \, \sum_b p_b^j \Big>\biggr] \,.
\end{align}
\end{widetext}
The large-$k$ expansion of the functions $P_{Li}(k)$ has the form
\begin{align}\label{eq:15}
k\,P_{Li}(k) &\ = G_i k^2 + F_i k + A_i + \frac{B_i}{\sqrt{k}} + \frac{C_i\,\ln k}{k} + \frac{D_i}{k} + \ldots\,,
\end{align}
where the first two coefficients are nonzero only for the $P_{L3}$ term (i.e., $G_1 = G_2 = F_1 = F_2 = 0$).

The finite parts of the corrections $\Delta E_{Li}(\Lambda)$ in Eq.~(\ref{eq:11}) are defined as
\begin{align}\label{eq:16}
E_{Li} = &\ \frac{2\alpha}{3\pi m^2} \int_0^{\infty} dk\,\biggl\{k\,P_{Li}(k) - k^2 G_i - k\,F_i
 - A_i
\nonumber \\ &
 - \frac{B_i}{\sqrt{k}} - \biggl[\frac{C_i\,\ln k}{k} + \frac{D_i}{k}\biggr] \theta(k-E_h)\biggr\}\,.
\end{align}
For the numerical evaluation, it is convenient to transform the above expression to an equivalent form,
\begin{align}\label{eq:17}
E_{Li}  &\ = \frac{2\alpha}{3\pi m^2}  \biggl\{ \int_0^K dk\,k\, P_{Li}(k) + \int_K^{\infty} dk\,
\biggl[ k P_{Li}(k)
 \nonumber \\ &
- G_i k^2 - F_i k - A_i - \frac{B_i}{k^{1/2}}  - \frac{C_i\ln k}{k}  - \frac{D_i}{k}\biggr]
 \nonumber \\ &
  -G_i \frac{K^3}{3} - F_i \frac{K^2}{2} -A_i K - 2B\sqrt{K_i}
 \nonumber \\ &
  -\frac{C_i}{2}\ln^2K - D_i\ln K\biggr\}\,,
\end{align}
where $K \ge E_h$ is a free parameter. One can easily show that the result does not depend on the choice of $K$.

\section{Regularization}
\label{sec:reg}

From now on, we will present formulas explicitly for the two-electron atom. We will also use the
short-hand notation $r \equiv r_{12}$.

For the numerical evaluation of the perturbations induced by the Breit Hamiltonian, it is advantageous
to transform formulas to a more regular form, which leads to a much better numerical convergence. For the perturbed
wave-function part of $P_{L1}$ we introduce the following (non-Hermitian) regularized Breit operator $H_{\rm Breit}'$
\begin{align}\label{eq:r:1}
H_{\rm Breit}' = &\ -\frac12 (E-V)^2 + \frac14 \nabla_1^2 \nabla_2^2 - \frac{Z}{4} \frac{\vec{r}_1}{r_1^3}\cdot\vec{\nabla}_1
 \nonumber \\ &
- \frac{Z}{4} \frac{\vec{r}_2}{r_2^3}\cdot\vec{\nabla}_2
-\frac{1}{2}\,p_1^i\,
\biggl(\frac{\delta^{ij}}{r}+\frac{r^i\,r^j}{r^3}\biggr)\,p_2^j\,.
\end{align}
It can be shown that for any trial function $|\phi\rbr$, the following identity holds
\begin{align}\label{eq:r:2}
H_{\rm Breit}|\phi\rbr = H_{\rm Breit}'|\phi\rbr + \bigl\{ H-E,Q \bigr\}|\phi\rbr\,,
\end{align}
where
\begin{align}\label{eq:r:3}
Q = - \frac14 \left( \frac{Z}{r_1} + \frac{Z}{r_2} - \frac{2}{r} \right)\,.
\end{align}
Using the identity (\ref{eq:r:2}), we transform the perturbed
wave-function part of $P_{L1}$ to a more regular form as follows
\begin{align}\label{eq:r:4}
P_{L1, \rm pwf}(k) = &\ 2 \Big< H_{\rm Breit}' \frac1{(E-H)'}\, \vec{P}\,  \frac1{E-H-k} \, \vec{P} \Big>
 \nonumber \\ &
 - 2 \Big< \bigl[ Q - \lbr Q\rbr \bigr] \, \vec{P}\,  \frac1{E-H-k} \, \vec{P} \Big>\,.
\end{align}

\begin{widetext}
For the vertex part of $P_{L1}$, we use a more complicated, Hermitian version of the regularized
Breit operator,
\begin{align}\label{eq:r:5}
  H_{\rm Breit}'' = -\frac12 (E-V)\,\left(E-\frac{1}{r}\right) + \frac14 \nabla_1^2 \nabla_2^2
  - \frac{Z}{4} \vec p_1\left(\frac{1}{r_1}+\frac{1}{r_2}\right)\vec p_1
  - \frac{Z}{4} \vec p_2\left(\frac{1}{r_1}+\frac{1}{r_2}\right)\vec p_2
-\frac{1}{2}\,p_1^i\,\biggl(\frac{\delta^{ij}}{r}+\frac{r^i\,r^j}{r^3}\biggr)\,p_2^j\,.
\end{align}
For this operator, the following identity holds
\begin{align}\label{eq:r:6}
H_{\rm Breit} = H_{\rm Breit}'' + \bigl\{ H-E,Q' \bigr\} -\frac{1}{2}\,(H-E)^2\,,
\end{align}
where $Q' = Q - \frac{E}{2}$. Using this identity, we derive the following regularized expression for
the vertex part of $P_{L1}$,
\begin{align}
P_{\rm ver}(k) = &\  \Big< \vec{P}\,  \frac1{E-H-k} \,\Bigl[ H_{\rm Breit}'' -2 k Q + kE -\frac{k^2}{2}- \lbr H_{\rm Breit}\rbr\Bigr]  \frac1{E-H-k} \,\vec{P} \Big>
\nonumber \\ &
  - \Big<\bigl[ 2\vec{P}\,Q+(k-E)\vec{P}\bigr]  \frac1{E-H-k}  \,\vec{P} \Big>  - \frac12 \Big< \vec{P}^2 \Big>\,.
\end{align}

\section{Angular reduction}

We now turn to performing the angular reduction of the above formulas for the $^3S$ and $^3P$ states of
a two-electron atom. The angular reduction is carried out in Cartesian coordinates. The representation of wave functions
in Cartesian coordinates is discussed in detail in Ref.~\cite{yerokhin:21:hereview}.

We start with the nonrelativistic Bethe logarithm.
The angular reduction of $P_{nd}$ for the $^3S$ reference state is trivial, since
only one angular symmetry ($^3P^o$) of intermediate states is allowed.
For the $^3P$ reference state, we decompose the Cartesian product of the current $j^i \equiv P^i$ and the
wave function $\phi^k$ into a sum of irreducible tensors of the rank $L = 0$, $1$, and $2$
as follows
\begin{align} \label{eq:a:1}
j^i \phi^k = &\ \frac13\,\delta^{ik} \vec{j}\cdot\vec{\phi} + \frac12\, \eps_{ikl} \bigl(
\vec{j}\times\vec{\phi}\bigr)_l
+ \frac12\biggl[j^i \phi^k + j^k \phi^i -\frac23 \delta^{ik} \vec{j}\cdot\vec{\phi}\biggr]\,.
\end{align}
This decomposition leads to the separation of $P_{nd}$ into the
contributions with $^3S$, $^3P^e$, and $^3D^e$ intermediate states,
\begin{align}\label{eq:a:2}
P_{nd}(k) &\ =
 \frac13\, \left< \phi^i \biggl| j^i \, \frac1{E-H-k}\biggr|_{^3S} \, j^k \biggr| \phi^k \right>
+ \frac12\, \left< {\Psi}^i_1  \, \frac1{E-H-k}\biggr|_{^3P^e} \,\Psi^i_1 \right>
+ \frac14\, \left< {\Psi}^{ik}_2  \, \frac1{E-H-k}\biggr|_{^3D^e} \,\Psi^{ik}_2 \right>
\,,
\end{align}
where $\vec{{\Psi}}_1 =  \vec{ j}\times\vec{\phi}$ and ${\Psi}_2^{ik} = j^i \phi^k + j^k \phi^i
-\frac23 \, \delta^{ik} \bigl( \vec{j}\cdot\vec{\phi}\bigr)$ and the summation over the repeated indices 
is implicit.

The angular reduction of $P_{L1}$ and $P_{L2}$ follows the same pattern as for the leading contribution $P_{nd}$.
For $P_{L3}$, we need first to perform the angular integration over $\hk$.
It is carried out with help of the following formulas
\begin{align}\label{eq:a:3}
\int \frac{d\hk}{4\pi} \, \hk^i \hk^j &\ = \frac13 \delta^{ij}\,,
\ \ \
\int \frac{d\hk}{4\pi} \, \hk^i \hk^j \hk^k \hk^l = \frac1{15} \left(\delta^{ij}\delta^{kl}
 + \delta^{il}\delta^{kj}+ \delta^{ik}\delta^{jl}\right) \,, \\
\int \frac{d\hk}{4\pi} \, \Big( \delta^{ij} - &\ \hk^i \hk^j\Big) \hk^n \hk^m r_1^nr_2^m  =
\frac1{15} \left(4\,\delta^{ij}(\bfr_1\cdot\bfr_2) - r_1^i r_2^j - r_1^jr_2^i \right)\,.
\end{align}
Performing the angular integration and using the fact that
$\vec L = \vec r_1\times\vec p_1 + \vec r_2\times\vec p_2$ is the angular
momentum operator commuting with $H$, we obtain
\begin{align}\label{eq:a:4}
P_{L3}(k) = &
\frac{k^2}{10} \Biggl[
3 \left< \bigl( p_1^{i} r_1^{j} + p_2^{i} r_2^{j}\bigr)^{(2)} \, \frac1{E-H-k} \,
         \bigl( r_1^{j} p_1^{i} + r_2^{j} p_2^{i}\bigr)^{(2)} \right>
  -\frac{5}{2\,k}\, \left< \vec L^2 \right>
         \nonumber \\ &\
  -2\, \left< \biggl[ p_1^i \bigl( 2\,\delta^{ij}r_1^2 - r_1^ir_1^j\bigr)
                    +p_2^i \bigl( 2\,\delta^{ij}r_2^2 - r_2^ir_2^j\bigr) \biggr] \, \frac1{E-H-k} \,
         \bigl( p_1^j + p_2^j\bigr) \right>\Biggr]\,,
\end{align}
where $(a^ib^j)^{(2)} = (a^ib^j + a^jb^i)/2 - (\vec a\cdot\vec b)\,\delta^{ij}/3$.
\end{widetext}
The angular reduction of the last term in Eq.~(\ref{eq:a:4}) is exactly the same as for $P_{nd}$,
$P_{L1}$ and $P_{L2}$. Let us now consider the angular reduction of the first term
Eq.~(\ref{eq:a:4}), which will be referred to as the symmetric part $P_{L3}^{\rm sym}$.
In the case of the $^3S$ reference state, there is a single angular-symmetry contribution of the $^3D^e$
type in the resolvent. The result reads
\begin{align}
P_{L3}^{\rm sym}(k) = \frac{3k^2}{40}\, \left< \Psi^{ik}_2  \, \frac1{E-H-k}\biggr|_{\,{^3D^e}} \,\Psi^{ik}_2 \right>\,,
\end{align}
where
\begin{align}
\bigl|{\Psi}_2^{ik}\bigr> = &\ \biggl( r_1^i p_1^k + r_1^k p_1^i  - \frac23\,\delta^{ik} \bfr_1\cdot\bfp_1
\nonumber \\ &
  + r_2^i p_2^k + r_2^k p_2^i
  - \frac23\,\delta^{ik} \bfr_2\cdot\bfp_2\biggr)
  \bigl|\phi\bigr>\,.
\end{align}

In order to perform the angular reduction of the symmetric part for the $^3P$ state, we use the following identity:
\begin{align} \label{30}
\frac12\sum_a \bigl( r_a^i p_a^j + r_a^jp_a^i\bigr)\,\phi^k
    = &\  T^{ijk} + \epsilon^{ikl}\,T^{lj} + \epsilon^{jkl}\,T^{li}
\nonumber \\ &
    +\delta^{ik}\,T^j + \delta^{jk}\,T^i + \delta^{ij}\,T'^k\,,
\end{align}
where $T^i$, $T^{ij}$, and $T^{ijk}$ are the irreducible Cartesian tensors of the first,
second, and third rank, respectively,
\begin{align}
    T^{ijk} \equiv& \ \sum_a (r_a^i\,p_a^j\,\phi^k)^{(3)}\,,\\
    T^{ij} =& \frac1{12}\,\sum_a \biggl[
    \epsilon^{jlm}\,\bigl(r_a^{i}\,p_a^{l}+r_a^{l}\,p_a^{i}\bigr)\,\phi^m
 \nonumber \\ &
  + \epsilon^{ilm}\,\bigl(r_a^{j}\,p_a^{l}+r_a^{l}\,p_a^{j}\bigr)\,\phi^m
    \biggr] \,, \\
    T^{i} =&  \frac1{20}\,\sum_a \biggl[
    3\,\bigl(r_a^{i}\,p_a^{l}+r_a^{l}\,p_a^{i}\bigr)\,\phi^l - 2\,r_a^l\,p_a^l\,\phi^i
    \biggr]\,, \\
    T'^{i} =& \frac{1}{10} \sum_a \biggl[ 4 r_a^l\,p_a^l\,\phi^i - r_a^{i}\,p_a^{l}\,\phi^l
    -  r_a^{l}\,p_a^{i}\,\phi^l\biggr]\,.
\end{align}
Every $T$ is a symmetric and traceless tensor. One does not need the explicit form of $T^{ijk}$
because when projected onto the state with $L=3$, it is automatically becomes irreducible,
so one can use the left side of Eq. (\ref{30}) instead.
As a check, all the terms except for the first one in the right-hand-side of Eq.~(\ref{30}) should
vanish when projected on the $L=3$ state.

The symmetric part is the sum of the $L = 1$, $2$, and $3$ parts, given by
\begin{align}
P_{L3}^{\rm sym}(k) =& \ \frac{3\,k^2}{2}\,
\Bigg[ \frac{4}{3}\,\Big<  {T^{\dag}}^{i}\,\left. \frac{1}{E-H-k}\right|_{\ ^3P^o}\,T^i\Big>
 \nonumber \\ &
+ \frac{6}{5}\,\Big<  {T^{\dag}}^{ij}\,\left.\frac{1}{E-H-k}\right|_{\ ^3D^o}\,T^{ij}\Big>
 \nonumber \\ &
+ \frac{1}{5}\,\Big<  {T^{\dag}}^{ijk}\,\left.\frac{1}{E-H-k}\right|_{\ ^3F^o}\,T^{ijk}\Big>
\Bigg]\,.
\end{align}

\section{Numerical evaluation}

For the numerical evaluation of the relativistic corrections to the Bethe logarithm
we need to be able to compute the integrands $P_{Li}(k)$ for different values of $k$
with a high precision. The crucial
part is to obtain highly accurate basis-set representations of the electron propagator 
$(E-H-k)^{-1}$ for various angular-momentum symmetries. 
The general idea is
to use the variational optimization of the basis for the cases when the integrand has
a form of a symmetric second-order perturbation correction, since then it obeys
the variational principle \cite{korobov:04}. Specifically, variational optimization can be
used for the nonrelativistic contribution $P_{nd}(k)$ and for the symmetric part of the
retardation contribution, $P_{L3}^{\rm sym}(k)$. These two cases cover
all angular-momentum symmetries in the electron propagator 
required in this work. Specifically, for the $^3S$ reference
state there are only two symmetries required ($^3P$ and $^3D$), whereas for the $^3P$ reference
state there are six different symmetries contributing to the final result. For each
angular-momentum symmetry, we perform a variational optimization of $P_{nd}(k)$ and
$P_{L3}^{\rm sym}(k)$ for four values of the photon momentum $k_i = (10^1,10^2,10^3,10^4)$.
The optimization was carried out with gradually increasing the size of the basis until
the convergence condition for the relative accuracy $\epsilon = 10^{-12}$
or the maximum size of the basis $N = 1400$ was reached. The optimized values of
nonlinear parameters were stored and then used for computation of $P_{Li}(k)$.

For a given value of $k$, the functions $P_{Li}(k)$ were computed with a basis obtained by merging together the
optimized sets for the two closest $k_i$ points, thus essentially
doubling the number of the basis functions. In this way, we were able to compute the functions $P_{L2}(k)$
for $k \le 10^4$ and $P_{L3}(k)$ for $k \le 10^3$ with 10-12 digits of accuracy. The calculation of $P_{L1}(k)$
is more complicated since it involves the Breit Hamiltonian, which remains quite singular even after the
regularization, so that
additional steps are needed. First, we compute and store the reference-state wave function perturbed
by the regularized Breit Hamiltonian $H_{\rm Breit}'$, $|\delta \psi\rbr = 1/(E-H)'|H_{\rm Breit}'\rbr$.
In order to get accurate results for the perturbed
wave function, we optimize basis for the symmetric second-order correction induced by $H_{\rm Breit}'$
and use this basis for calculating the perturbed wave function.
The convergence of results is rather slow, which is due to the fact that the perturbed wave function
$|\delta \psi\rbr$ has an integrable singularity at $r_a\to 0$. In
order to represent such wave functions with the exponential basis, very large (both positive and
negative) values of nonlinear parameters were required. In order to effectively span large regions of parameters, we
used non-uniform distributions, see Ref.~\cite{yerokhin:21:hereview}
for details. In actual calculations, we performed the variational optimization gradually increasing
the basis size up to $N = 1200$ and then doubled the basis when computing the perturbed wave function.
For other electron propagators in $P_{L1}(k)$ we used the same
numerical procedure as for $P_{L2}(k)$ and $P_{L3}(k)$. In this way were able to compute
the function $P_{L1}(k)$ for $k \le 10^4$ with accuracy of about 9 digits.

The final step is the computation of the relativistic corrections $E_{Li}$ according to
Eq.~(\ref{eq:17}). The interval of the photon momenta $k \in (0,\infty)$ is split in two by the parameter $K$. In
this work we use $K = 100$. The integral over the interval $(0,K)$ is carried out
analytically, by diagonalizing the Hamiltonian matrix and using the spectral representation of the
electron propagator. We note that the principal value of the integral should be taken when the
intermediate-state energies smaller than the reference-state energy occur.
In this way the integral over $(0,K)$  is evaluated without any loss of numerical precision.  
The second part of the integral over $(K,\infty)$ is
evaluated by integrating the large-$k$ expansion of the integrand, with the coefficients of
the expansion obtained by fitting the numerical values of the integrand to the known form
of the asymptotic expansion.

For $P_{L2}$ and $P_{L3}$, we use the large-$k$ expansion of the form \cite{korobov:13}
\begin{align}\label{eq:num1}
k P_{Li}(k) -k^2G_i &\  - k F_i - A_i - \frac{B_i}{\sqrt{k}}  - \frac{C_i\ln k}{k}  - \frac{D_i}{k}
 \nonumber \\ &
= \frac1k\,\sum_{m = 1}^M \frac{c_{m,2} \sqrt{k} + c_{m,1}\ln k + c_{m,0}}{k^{m}}\,,
\end{align}
where the coefficients $c_{m,n}$ are obtained from the fitting procedure.
The large-$k$ expansion of $P_{L1}$ is more complicated, \cite{korobov:13}
\begin{align}\label{eq:num2}
k P_{L1}(k) - A_1 - \frac{B_1}{\sqrt{k}} &\ - \frac{C_1\ln k}{k}  - \frac{D_1}{k}
 \nonumber \\ &
 =
\frac1k\,\sum_{m = 1}^M \sum_{n = 0}^m \frac{c_{m,n} \ln^n k}{k^{m/2}}\,,
\end{align}
with coefficients $c_{m,n}$ to be determined numerically. The coefficients $G_i$, $F_i$, $A_i$, $B_i$, $C_i$, and
$D_i$ are known analytically; explicit formulas them are presented in Ref.~\cite{yerokhin:18:betherel}.
Note that the definition of $P_{L2}$ in this work (and, therefore, definitions of the corresponding asymptotic constants)
differ from Ref.~\cite{yerokhin:18:betherel} by a factor of 2.

The fitting was performed as follows.
At the first step, we store numerical values of the functions $P_{Li}(k)$ for different values of $k$
in the interval $k \in (5,10^4)$ (typically, about 300 points).  For $P_{L3}(k)$,
numerical cancellations  in subtracting the large-$k$ asymptotics are larger, so we used
a smaller interval $k \in (5,10^3)$.
At the second step, we subtract contributions of all asymptotic constants known analytically
except $D_i$ from the stored values and select several variants of fitting functions and
fitting intervals $k \in (k_{\rm min}, k_{\rm max})$
that yield the best results for the asymptotic constant $D_i$.
Typically, 10-16 free parameters in the fitting anzatz were used.
Finally, we use the analytical results
for $D_i$ and apply the optimal fitting prescriptions to obtain results for the high-$k$ part
of the integral. The scattering of results obtained with different fitting functions were used
for estimating the uncertainty.

%

\begin{table}
\caption{Relativistic corrections to the Bethe logarithm for the $2\,^3S$ state.
\label{tab:S}}
\begin{ruledtabular}
\begin{tabular}{lw{1.5}w{2.5}w{2.5}w{2.8}c}
\multicolumn{1}{l}{$Z$}
        & \multicolumn{1}{c}{$\beta_{L1}$}
            & \multicolumn{1}{c}{$\beta_{L2}$}
                & \multicolumn{1}{c}{$\beta_{L3}$}
                & \multicolumn{1}{c}{$\beta_L$}
                & \multicolumn{1}{c}{Ref.}
  \\
\hline\\[-5pt]
   2 &  -3.335\,97  &  16.963\,35  & -40.596\,75 & -26.969\,37\,(2) \\
     &  -3.335\,96  &  16.963\,47  & -40.596\,75 & -26.969\,2\,(2) & \cite{yerokhin:18:betherel}\\
   3 &  -3.396\,19  &  16.917\,55  & -40.584\,84 & -27.063\,48\,(2) \\
   4 &  -3.433\,87  &  16.884\,51  & -40.578\,19 & -27.127\,56\,(2) \\
   5 &  -3.458\,95  &  16.861\,90  & -40.574\,00 & -27.171\,05\,(2) \\
   6 &  -3.476\,71  &  16.845\,75  & -40.571\,12 & -27.202\,08\,(2) \\
   7 &  -3.489\,91  &  16.833\,71  & -40.569\,04 & -27.225\,24\,(2) \\
   8 &  -3.500\,09  &  16.824\,41  & -40.567\,46 & -27.243\,14\,(2) \\
   9 &  -3.508\,17  &  16.817\,02  & -40.566\,23 & -27.257\,38\,(3) \\
  10 &  -3.514\,76  &  16.811\,02  & -40.565\,23 & -27.268\,98\,(4) \\
  11 &  -3.520\,21  &  16.806\,04  & -40.564\,42 & -27.278\,59\,(4) \\
  12 &  -3.524\,88  &  16.801\,85  & -40.563\,74 & -27.286\,77\,(5) \\
 $\infty$ &                &              &      & -27.381\,4\,(6)\\
          &               &              &      & -27.381\,138   & \cite{jentschura:05:sese}\\
\end{tabular}
\end{ruledtabular}
\end{table}

\begin{table}
\caption{Relativistic corrections to the Bethe logarithm for the $2\,^3P$ state.
\label{tab:P}}
\begin{ruledtabular}
\begin{tabular}{lw{1.5}w{2.5}w{2.5}w{2.8}c}
\multicolumn{1}{l}{$Z$}
        & \multicolumn{1}{c}{$\beta_{L1}$}
            & \multicolumn{1}{c}{$\beta_{L2}$}
                & \multicolumn{1}{c}{$\beta_{L3}$}
                & \multicolumn{1}{c}{$\beta_L$}
                & \multicolumn{1}{c}{Ref.}
  \\
\hline\\[-5pt]
   2 &  -3.292\,74   &  16.939\,85                 & -40.644\,79                 & -26.997\,68\,(25) \\
     &  -3.292\,77   &  16.939\,94                 & -40.644\,78                 & -26.997\,6\,(5)  & \cite{yerokhin:18:betherel}\\
   3 &  -3.276\,93   &  16.915\,94                 & -40.678\,95                 & -27.039\,94\,(20) \\
   4 &  -3.267\,96   &  16.884\,90                 & -40.702\,03                 & -27.085\,08\,(20) \\
   5 &  -3.264\,40   &  16.861\,08                 & -40.717\,78                 & -27.121\,09\,(20) \\
   6 &  -3.263\,22   &  16.843\,21                 & -40.729\,10                 & -27.149\,11\,(20) \\
   7 &  -3.263\,01   &  16.829\,52                 & -40.737\,61                 & -27.171\,10\,(20) \\
   8 &  -3.263\,27   &  16.818\,77                 & -40.744\,24                 & -27.188\,74\,(20) \\
   9 &  -3.263\,80   &  16.810\,12                 & -40.749\,54                 & -27.203\,21\,(20) \\
  10 &  -3.264\,36   &  16.803\,03                 & -40.753\,88                 & -27.215\,20\,(20) \\
  11 &  -3.264\,92   &  16.797\,12                 & -40.757\,49                 & -27.225\,29\,(20) \\
  12 &  -3.265\,44   &  16.792\,11                 & -40.760\,55                 & -27.233\,88\,(20) \\
 $\infty$ &               &                             &                        & -27.340\,8\,(30)\\
          &   &             &                                                   & -27.341\,771  & \cite{jentschura:03:prl,jentschura:05:sese}\\
\end{tabular}
\end{ruledtabular}
\end{table}

%

\section{Results}

The relativistic corrections to the Bethe logarithm were calculated for the helium atom
in Ref.~\cite{yerokhin:18:betherel}, defined as given by Eq.~(\ref{eq:16}). For helium-like
ions, however, this definition is not very convenient. The reason is that the $Z$ dependence
of the corrections $E_{Li}$ is quite complicated; they scale as $Z^6$ and in addition contain terms proportional
to $\ln Z$ and $\ln^2Z$. It is thus advantageous to separate out the leading $Z$ dependence and
logarithmic terms from the
definition, similarly to that for the nonrelativistic Bethe logarithm (\ref{eq:5}). The separation
of logarithms can be achieved
by changing the cutoff parameter in Eq.~(\ref{eq:16}), $E_h = m\alpha^2 \to m(\Za)^2$.

So, instead of corrections $E_{Li}$ we introduce the
functions $\beta_{Li}$ that do not have
logarithmic terms in their $1/Z$ expansion and are related to $E_{Li}$ as follows
\begin{align}
\beta_{Li} = &\ \frac1{Z^3 \lbr \sum_a\delta^3(r_a) \rbr} \Big[ E_{Li}
 + \frac{2}{3\pi} \Big( \frac{C_i}{2}\,\ln^2Z^2 + D_i \ln Z^2\Big)
 \Big]\,,
\end{align}
where $C_i$ and $D_i$ are the large-$k$ asymptotic expansion constants in Eq.~(\ref{eq:16}), explicit
formulas for which can be found in Ref.~\cite{yerokhin:18:betherel}.

In the high-$Z$ limit, the functions $\beta_{Li}(Z)$  
should approach the asymptotic values that can be obtained from the hydrogen theory. Specifically,
for a two-electron $1snl$ state the large-$Z$ limit is obtained as
\begin{align}
\beta_{0}(1snl) = \Big( 1 + \frac{\delta_{l,0}}{n^3}\Big)^{-1} \, \Big[ {\cal L}(1s) + \frac{{\cal L}(nl)}{n^3} \Big]\,,
\end{align}
where ${\cal L}(nl)$ is the one-loop hydrogenic low-energy contribution from Refs.~\cite{jentschura:03:prl,jentschura:05:sese},
${\cal L}(np) = (1/3)\, {\cal L}(np_{1/2})+ (2/3)\, {\cal L}(np_{3/2})$.

\begin{figure*}
\centerline{
\resizebox{.95\textwidth}{!}{%
  \includegraphics{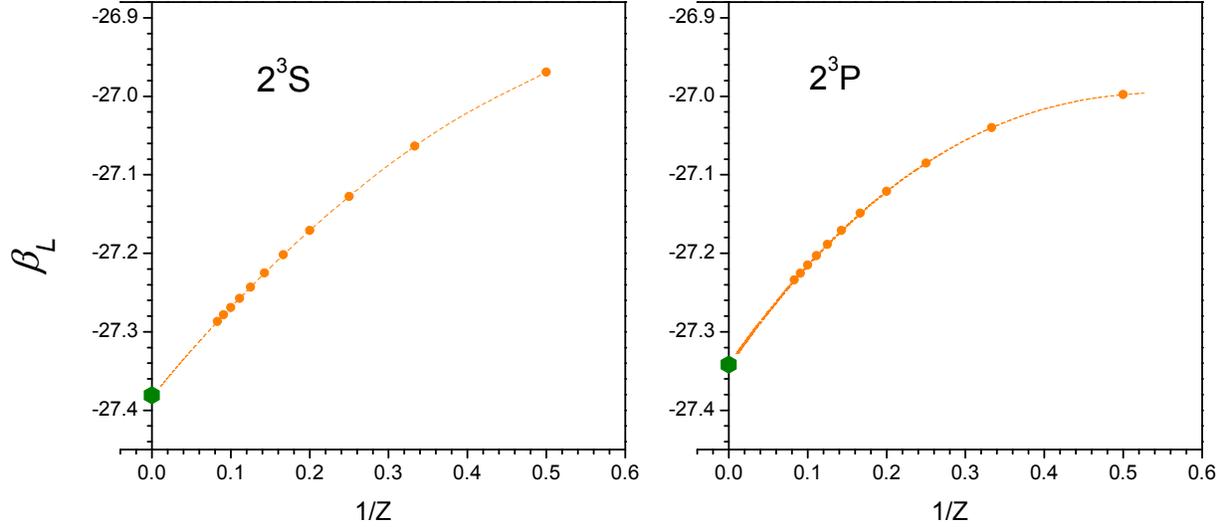}
}}
 \caption{The relativistic correction to the Bethe logarithm $\beta_L$ for the $2\,^3S$ (left) and $2\,^3P$ (right) states
 of helium-like ions, as a function of the inverse nuclear charge $1/Z$. Round dots (orange) denote the numerical results,
 the hexagon dot (green) shows the analytical result at $Z= \infty$, dotted line (orange) represents the numerical fit.
 \label{fig:1}}
\end{figure*}

Results of our numerical calculations of the relativistic corrections to the Bethe logarithm for the
$2\,^3S$ and $2\,^3P$ states of helium-like atoms with $Z \le 12$ are collected in Tables~\ref{tab:S}
and \ref{tab:P} and Fig.~\ref{fig:1}. We observe that the numerical values of $\beta_L$ exhibit a weak dependence on
$Z$. Moreover, both for the $2\,^3S$ and $2\,^3P$ states the results are quite close to the hydrogenic $1s$
value ${\cal L}(1s) = -27.259\,909$ \cite{jentschura:05:sese}. This behaviour is similar to that of the
nonrelativistic Bethe logarithm \cite{drake:99:cjp}.

Tables~\ref{tab:S}
and \ref{tab:P} present also results of our numerical extrapolation of $\beta_L(Z)$ to the $Z\to \infty$ limit.
The extrapolation was carried out by fitting the $1/Z$ expansion to a polynomial
in $1/Z$. We observe that the fitting results are in excellent agreement with the analytical values of
$\beta_0$ obtained from the hydrogen theory \cite{jentschura:03:prl,jentschura:05:sese}.

Summarizing, we performed calculations of the relativistic corrections to the Bethe logarithm for the $2\,^3S$
and $2\,^3P$ states of helium-like ions with $Z \leq 12$. The leading $Z$ dependence and terms proportional
to $\ln Z$ and $\ln^2 Z$ were separated out. The resulting scaled function $\beta_L$ was found to depend weakly
on $Z$ and on the reference state. The extrapolated $Z\to \infty$ limit of the numerical results was
found to be in excellent agreement with the analytical values obtained from the hydrogen theory. This constitutes
a stringent check of correctness of the numerical procedure of the calculation.

\begin{acknowledgments}
The work was supported by the Russian Science Foundation (Grant No. 20-62-46006).
K.P. acknowledges support from the National Science Center (Poland) Grant No. 2017/27/B/ST2/02459.
Computations were performed partly in the computer cluster ``Tornado'' of St.~Petersburg Polytechnic University.
\end{acknowledgments}

\end{document}